\documentclass[twoside,twocolumn,9pt]{article}
\usepackage{extsizes}
\usepackage[super,sort&compress,comma]{natbib} 
\usepackage[version=3]{mhchem}
\usepackage[left=1.5cm, right=1.5cm, top=1.785cm, bottom=2.0cm]{geometry}
\usepackage{balance}
\usepackage{mathptmx}
\usepackage{sectsty}
\usepackage{graphicx} 
\usepackage{lastpage}
\usepackage[format=plain,justification=justified,singlelinecheck=false,font={stretch=1.125,small,sf},labelfont=bf,labelsep=space]{caption}
\usepackage{float}
\usepackage{fancyhdr}
\usepackage{fnpos}
\usepackage[english]{babel}
\addto{\captionsenglish}{%
  
}
\usepackage{array}
\usepackage{droidsans}
\usepackage{charter}
\usepackage[T1]{fontenc}
\usepackage[usenames,dvipsnames]{xcolor}
\usepackage{setspace}
\usepackage[compact]{titlesec}
\usepackage{hyperref}

\usepackage{epstopdf}

\definecolor{cream}{RGB}{222,217,201}

\begin{document}

\pagestyle{fancy}
\thispagestyle{plain}
\fancypagestyle{plain}{
\renewcommand{\headrulewidth}{0pt}
}

\makeFNbottom
\makeatletter
\renewcommand\LARGE{\@setfontsize\LARGE{15pt}{17}}
\renewcommand\Large{\@setfontsize\Large{12pt}{14}}
\renewcommand\large{\@setfontsize\large{10pt}{12}}
\renewcommand\footnotesize{\@setfontsize\footnotesize{7pt}{10}}
\makeatother

\renewcommand{\thefootnote}{\fnsymbol{footnote}}
\renewcommand\footnoterule{\vspace*{1pt}%
\color{cream}\hrule width 3.5in height 0.4pt \color{black}\vspace*{5pt}} 
\setcounter{secnumdepth}{5}

\makeatletter 
\renewcommand\@biblabel[1]{#1}            
\renewcommand\@makefntext[1]%
{\noindent\makebox[0pt][r]{\@thefnmark\,}#1}
\makeatother 
\renewcommand{\figurename}{\small{Fig.}~}
\sectionfont{\sffamily\Large}
\subsectionfont{\normalsize}
\subsubsectionfont{\bf}
\setstretch{1.125} 
\setlength{\skip\footins}{0.8cm}
\setlength{\footnotesep}{0.25cm}
\setlength{\jot}{10pt}
\titlespacing*{\section}{0pt}{4pt}{4pt}
\titlespacing*{\subsection}{0pt}{15pt}{1pt}

\fancyfoot{}
\fancyfoot[LO,RE]{\vspace{-7.1pt}\includegraphics[height=9pt]{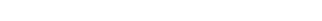}}
\fancyfoot[CO]{\vspace{-7.1pt}\hspace{11.9cm}\includegraphics{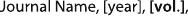}}
\fancyfoot[CE]{\vspace{-7.2pt}\hspace{-13.2cm}\includegraphics{head_foot/RF}}
\fancyfoot[RO]{\footnotesize{\sffamily{1--\pageref{LastPage} ~\textbar  \hspace{2pt}\thepage}}}
\fancyfoot[LE]{\footnotesize{\sffamily{\thepage~\textbar\hspace{4.65cm} 1--\pageref{LastPage}}}}
\fancyhead{}
\renewcommand{\headrulewidth}{0pt} 
\renewcommand{\footrulewidth}{0pt}
\setlength{\arrayrulewidth}{1pt}
\setlength{\columnsep}{6.5mm}
\setlength\bibsep{1pt}

\makeatletter 
\newlength{\figrulesep} 
\setlength{\figrulesep}{0.5\textfloatsep} 

\newcommand{\topfigrule}{\vspace*{-1pt}%
\noindent{\color{cream}\rule[-\figrulesep]{\columnwidth}{1.5pt}} }

\newcommand{\botfigrule}{\vspace*{-2pt}%
\noindent{\color{cream}\rule[\figrulesep]{\columnwidth}{1.5pt}} }

\newcommand{\dblfigrule}{\vspace*{-1pt}%
\noindent{\color{cream}\rule[-\figrulesep]{\textwidth}{1.5pt}} }

\makeatother

\twocolumn[
  \begin{@twocolumnfalse}
{\includegraphics[height=30pt]{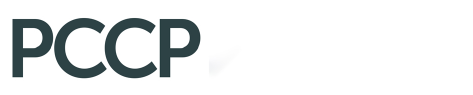}\hfill\raisebox{0pt}[0pt][0pt]{\includegraphics[height=55pt]{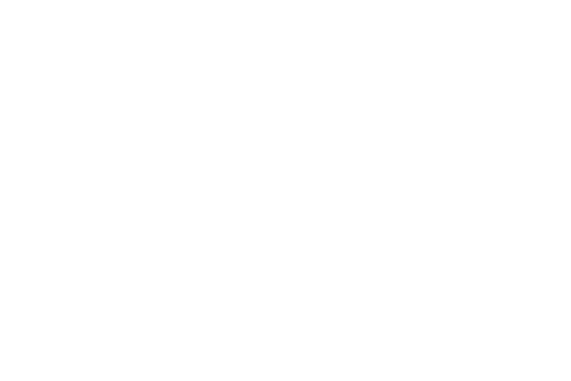}}\\[1ex]
\includegraphics[width=18.5cm]{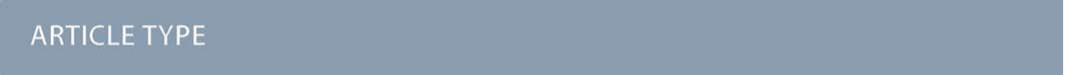}}\par
\vspace{1em}
\sffamily
\begin{tabular}{m{4.5cm} p{13.5cm} }

\includegraphics{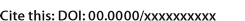} & \noindent\LARGE{\textbf{Production of ultracold asymmetric tops from Sr atoms and SrOH molecules}} \\
\vspace{0.3cm} & \vspace{0.3cm} \\

 & \noindent\large{Maciej B. Kosicki\textit{$^{a}\dag$}, Mateusz Borkowski\textit{$^{b}$},  Marcin Umi\'nski\textit{$^{b}$},  and Piotr S. \.Zuchowski\textit{$^{b}\ddag$}} \\
 

\includegraphics{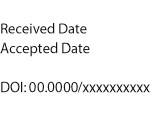} & \noindent\normalsize{We report the comprehensive theoretical investigation of the Sr–SrOH system identifying it as a promising route to production of ultracold asymmetric top molecules. 
Combining high-level \textit{ab initio} electronic structure calculations with rigorous quantum scattering simulations, we determine strongly anisotropic, non-reactive interaction potential and an exceptionally dense spectrum of near-threshold resonances. Presented excited states,  transition dipole moments, and results of the one-dimensional STIRAP model suggest the possibility of coherently transferring weakly bound complexes to the rovibrational ground state.} \\

\end{tabular}

 \end{@twocolumnfalse} \vspace{0.6cm}

  ]

\renewcommand*\rmdefault{bch}\normalfont\upshape
\rmfamily
\section*{}
\vspace{-1cm}

\footnotetext{\textit{$^{a}$~Faculty of Physics, Kazimierz Wielki University, al. Powsta\'nc\'ow Wielkopolskich 2, 85-090 Bydgoszcz, Poland.}}
\footnotetext{\textit{$^{b}$~Institute of Physics, Faculty of Physics, Astronomy and Informatics, Nicolaus Copernicus University,  Grudzi{\c a}dzka 5, 87-100 Toru\'n, Poland.}}
\footnotetext{$\dag$~E-mail: mac.kosicki@gmail.com}
\footnotetext{$\ddag$~E-mail: pzuch@fizyka.umk.pl}




\section{Introduction}

Ultracold polyatomic molecules have recently emerged as powerful platforms for probing physics beyond the current state of knowledge~\cite{Augenbraun_2023}. 
Their rich internal structure,  featuring opposite-parity states separated by a small energy gap, large permanent dipole moments, and rotational and vibrational degrees of freedom, makes them ideal candidates for the exploration of new physics~\cite{Hutzler_2020, Doyle_2022}. 
The long coherence times of trapped molecules allow precision spectroscopy of multi‑mode vibrational transitions that exhibit sensitivity to temporal variations in the proton‑to‑electron mass ratio \cite{Kozyryev_2021}. 
Long-lived bending vibrational levels can be fully polarized with fields and photon cycled, amplifying the sensitivity of searches for electron electric dipole moment~\cite{Kozyryev_2017, Anderegg_2023}, while symmetric top molecules leverage this capability to probe T-violating interactions\cite{Norrgard_2019, Augenbraun_2021, Yu_2021}.
Extending these considerations to asymmetric top molecules is particularly compelling. Because of their three unequal principal moments of inertia, these molecules lack the simplifying symmetries present in linear or symmetric top species and therefore display more complex rotational and vibrational structures. This structural richness introduces qualitatively new features, including independently controllable permanent dipole components, which can significantly advance both precision measurement and quantum science applications~\cite{Doyle_2022}.
In parallel, the combination of strong dipolar interactions and complex internal structure makes polyatomic species highly promising for quantum simulation~\cite{Wall_2013, Wall_2015} and quantum information processing~\cite{Wei_2011, Yu_2019, Albert_2020}.

To realize these opportunities, a particularly promising route relies on polyatomic molecules in the $X^2\Sigma^+$ electronic ground state, such as closed-shell ligand monohydroxides\cite{Isaev_2016, Kozyryev_2016, Augenbraun_2020}.
Their single valence electron resides in a metal‑centered orbital $s\sigma$, allowing nearly closed optical transitions directly analogous to the D lines in alkali-metal atoms and alkaline-earth metal atoms.
Vibrational branching is strongly suppressed by the near-diagonal Franck–Condon matrix and the similarity between ground- and excited-state potential energy surfaces.
The parity and angular momentum selection rules then permit fully closed rotational cycling schemes, while ligand‑induced orbital hybridization in monovalent alkaline-earth compounds minimizes losses due to vibrational excitation.
Combined, these features allow for the photon scattering required for beam slowing, magneto-optical trapping, and sub-Doppler cooling of complex species such as linear and symmetric top molecules\cite{Kozyryev_2017b, Kozyryev_2019, Zhang_2021}.  
Subsequent precision measurements of vibronic branching ratios in SrOH and CaOH have demonstrated the feasibility of scattering photons for efficient laser cooling and trapping ~\cite{Lasner_2022,Baum_2021}. 
The successful demonstration of magneto-optical traps, first realized for diatomic monofluorides~\cite{Barry_2014, Norrgard_2016, McCarron_2018, Truppe_2018, Williams_2018}, and subsequently for monohydroxides~\cite{Baum_2020, Vilas_2022, Lasner_2025}, establishes the foundation for future trapping of asymmetric top species.
At the same time, other efforts have extended the Doppler and Sisyphus techniques to YbOH~\cite{Augenbraun_2020b}, implemented Sisyphus cooling for H${}_2$CO~\cite{Prehn_2016} and CaOCH${}_3$~\cite{Mitra_2020}, collectively enriching the toolkit to reach the ultracold regime in polyatomic molecules.

Guided by theoretical insights, the range of polyatomic molecules considered for laser cooling has not only expanded but also revealed entirely new classes of candidates with properties tailored for precision control~\cite{Isaev_2016, Li_2019, Li_2020}.
More recently, a diverse set of species has been proposed for laser cooling and optical cycling transitions. In particular, radium monohydroxide has been highlighted as the heaviest polyatomic molecule amenable to direct laser cooling, offering both efficient photon cycling and sensitivity to violations of fundamental symmetries in physics~\cite{Isaev_2017, Zhang_2023}. 
The \textit{ab initio} calculations have further identified tetratomic species CaCCH and SrCCH as promising candidates, showing highly diagonal Franck–Condon factors (FCFs), short radiative lifetimes, and Doppler limits in the $\mu$K regime~\cite{Xia_2021}. 
Predictions now extend to aromatic-ligand complexes containing alkaline-earth optical cycling centers, ranging from benzene, phenol, and cyclopentadienyl to naphthalene, coronene, and even fullerenes, as well as extended frameworks incorporating alkynyl chains~\cite{Ivanov_2020, Zhu_2022, Dickerson_2021, Mitra_2022}. 
Furthermore, theoretical characterizations of lanthanide monohydroxides have reported their electronic states, static dipole moments, and zero‑field splittings, suggesting new platforms for symmetry violation experiments~\cite{Klos_2025}.

It should be mentioned in this context, that studies on cotrapped ultracold atoms and diatomic molecules have a well-established history, greatly advancing our understanding of reactive collisions~\cite{Meyer_2011, Quemener_2016}, reaction control~\cite{Son_2022}, and Feshbach resonances~\cite{Park_2023b, Morita_2024}. 
Pioneering works with a quantum gas of ultracold polar KRb molecules in their absolute ground state demonstrated coherent control over internal molecular structure, opening the way to ultracold chemistry~\cite{Ni_2008, Ospelkaus_2008,Ospelkaus_2010, deSilva_2025,Ni_2010, deMiranda_2011}. The realization of three‑dimensional optical lattices loaded with KRb molecules enabled studies of dipolar spin dynamics and bimolecular reaction pathways under controlled conditions~\cite{Yan_2013, Hu_2019}, while manipulation of the reactant spins using magnetic fields provided an additional tool for steering the reaction channels~\cite{Hu_2021}. 
Furthermore, complete state‑resolved measurements of reaction products in ultracold KRb–KRb collisions provided rigorous tests of statistical models for internal energy distribution~\cite{Liu_2021}, and the first direct observation of inter‑species spin‑dependent hyperfine interactions in atom–molecule collisions established a new benchmark to understand spin–rotation coupling at ultralow temperatures~\cite{Liu_2025}. 
One should also add that numerous theoretical studies have examined ultracold collisions of diatomic molecules with atoms and other diatomic molecules~\cite{Balakrishnan_1998, Forrey_1999, Balakrishnan_2000, Bohn_2000, Zhu_2001, Bodo_2002, Stoecklin_2002, Volpi_2002,  Balakrishnan_2003, Krems_2003, Stoecklin_2003, Soldan_2004, Lee_2005, Tscherbul_2006a, Tscherbul_2006b, Lara_2007, Tacconi_2007, Hutson_2009, Tscherbul_2009, Soldan_2009, Skomorowski_2011, Skomorowski_2011a, Tokunaga_2011, Tscherbul_2011b, Zuchowski_2011, Feng_2012, Tscherbul_2012, Gonzalez_Martinez_2013a, Gonzalez_Martinez_2013b, Lim_2015, Morita_2018, daSilva_2023, Morita_2024}. 
Their insights have substantially deepened the interpretation of the experimental data and guided the design of new measurements~\cite{Hummon_2008, Tscherbul_2010, Hummon_2011, Son_2020, Jurgilas_2021, Park_2023b}. However, cotrapping of polyatomic molecules with atoms remains a major experimental challenge and has not yet been achieved. There are also very few theoretical studies on such systems in ultracold regime~\cite{Zuchowski_2008, Zuchowski_2009, Tscherbul_2011, Morita_2017}.  This paper will shed some light on this problem.

One of the most promising pathways for exploring ultracold matter is to bring two different species into contact—either to thermalize them, to create new composite species such as molecules, or to initiate chemical reactions in the ultracold regime. The chemistry of polyatomic ultracold molecules is largely unexplored, with only the exception of Li-SrOH collisions in magnetic field~\cite{Morita_2017}. 
In this work, we investigate the production of ultracold asymmetric top molecules by combining strontium atoms with a polyatomic radical SrOH, focusing on their electronic structure, intermolecular interactions, cold collisions, and prospects for coherent molecule formation using the simplified concept of stimulated Raman adiabatic passage (STIRAP). 
Using high-level \textit{ab initio} methods, we determine for the first time the static polarizability tensor of SrOH, analyze its interactions with a family of $s^2$-type atoms (Mg, Ca, Sr, Yb, Hg), and map the detailed potential energy surface (PES) of the Sr–SrOH complex in its $X^2\Sigma^+$ electronic ground state.
Our calculations reveal that Sr–SrOH is a non-reactive system under ultracold conditions and features a strongly anisotropic PES. This anisotropy gives rise to a dense spectrum of near-threshold bound states, leading to an exceptionally rich set of scattering resonances. 
Finally, by characterizing low-lying electronically excited states of Sr$_2$OH and their transition dipole moments, we identify specific channels that may enable two-step STIRAP transfer from weakly bound complexes to the rovibrational ground state. 

This paper is organized as follows.
In Section~\ref{sec:comp_details}, we describe the \textit{ab initio} electronic structure methods used to compute the electric properties of SrOH, optimize the geometries of its complexes with alkaline-earth metal atoms, and implement potential energy surfaces for the Sr–SrOH interaction. 
We also outline the quantum scattering calculations including the treatment of anisotropic couplings and potential energy scaling to probe resonance sensitivity.
Sections~\ref{sec:results_properties}-\ref{sec:results_pathway} present our numerical results covering the permanent dipole moment and static polarizability of SrOH, systematic trends in s$^2$ metal–SrOH geometry, the topology and anisotropy of the Sr–SrOH PES, resonance structures in cold collisions, as well as excited-state potential curves and transition dipole moments relevant to coherent optical transfer, including a pathway for forming ground-state Sr$_2$OH molecules.
In conclusion, Section~\ref{sec:summary} summarizes our main findings and discusses their implications for ultracold asymmetric top formation.

\section{Computational Details}
\label{sec:comp_details}

\subsection{Electronic structure of the ground state}
\label{sec:comp_details_ground_state}

The electronic structure was studied using high‐quality, correlation-consistent basis sets and relativistic pseudopotentials, as appropriate for each element. 
For lighter atoms, we employed correlation-consistent basis sets augmented with diffuse functions.
Specifically, hydrogen and oxygen were described using the aug-cc-pVTZ basis set ~\cite{Kendall_1992a}. 
Calcium and magnesium were treated with the Douglas–Kroll–Hess recontracted aug-cc-pVTZ-DK basis sets~\cite{Prascher_2011a, Hill_2017}. 
Strontium was modeled with the small-core ECP28MDF pseudopotential and its corresponding valence \textit{spdf} basis set~\cite{Lim_2005, Lim_2006}. 
Ytterbium and mercury were described using the ECP60MDF energy-consistent pseudopotential, in combination with an uncontracted valence basis set reduced to \textit{spdf} functions~\cite{Wang_1998} for ytterbium, and the aug-cc-pVQZ-PP basis set including \textit{spdf} functions with optimized contractions for mercury~\cite{Figgen_2005, Peterson_2005}.
All electronic structure calculations were carried out using \textsc{molpro}\cite{MOLPRO_brief:2015}.

Geometries of the electronic ground state ($X^2\Sigma^+$) for the $\text{Mg}-$$\text{SrOH}$, $\text{Ca}-$$\text{SrOH}$, $\text{Sr}-$$\text{SrOH}$, and $\text{Yb}-$$\text{SrOH}$ complexes were optimized using the spin‑restricted coupled cluster approach with single, double, and perturbative triple excitations [RCCSD(T)], employing a restricted Hartree–Fock (RHF) reference wavefunction. Geometry optimization for the $\text{Hg}-$$\text{SrOH}$ complex was carried out at the Restricted M{\o}ller--Plesset Perturbation Theory of Second Order level (RMP2)~\cite{Hampel_1992, Deegan_1994}.  Based on these optimized geometries, the RCCSD(T)~\cite{Watts_1993, Knowles_1993} interaction energy was computed using the supermolecular approach with counterpoise correction to eliminate basis set superposition error (BSSE)\cite{Boys_1970}. The permanent dipole moments were then evaluated at the multi-configurational self-consistent Field (MCSCF) level by computing the expectation value of the electric‑dipole operator~\cite{Kreplin_2020}.

We also investigated the response properties of SrOH using the finite-field approach. For this purpose, a uniform dipole field with a step size of 0.001 a.u. was applied to the one-electron Hamiltonian, and finite difference derivatives were evaluated to extract both the electric dipole moment and the diagonal components of the static polarizability tensor at the RCCSD(T) level.
To quantify the effect of electron correlation, we repeated the same calculations at the RHF level and directly compared the resulting properties.

We studied the relative energetic stability of three dissociation channels for the linear Sr-SrOH complex in the ground state.
In the SrOH + Sr asymptotic channel, the Sr–O and O–H bond lengths were fixed at 2.15904 $\text{\AA}$~\cite{Li_2020} and 0.96944 $\text{\AA}$~\cite{Li_2020}, for the SrO + SrH asymptotic channel, Sr–O and Sr–H distances were fixed at  1.920 $\text{\AA}$~\cite{Field_1974} and 2.1456 $\text{\AA}$~\cite{Huber_1979}, and in the Sr${}_2$ + OH asymptotic channel, Sr–Sr and O–H bond lengths of 1.920 $\text{\AA}$~\cite{Leung_2021} and 0.970 $\text{\AA}$~\cite{Huber_1979} were used. 
For all three channels, the distances between the monomers were set at 100.0 $\text{\AA}$.
We adopted the binding energy of the Sr-SrOH complex as a zero-energy reference. 
Relative energies for each dissociation channel were obtained by subtracting this reference energy from the corresponding total energy.

The selected cuts through the potential energy surfaces were calculated in Jacobi coordinates, defined by the Jacobi distance $\textbf{R}$ between the metal atom (Mg, Ca, Sr, Yb, or Hg) and the center of mass of the rigid SrOH molecule, and by the Jacobi angle $\theta$ between the metal–COM vector and the molecular axis, at the RCCSD(T) level of theory. The Sr–O and O–H bond lengths in the linear SrOH molecule were taken from Li et al.\cite{Li_2020} 
The radial coordinate was represented on a grid of 100 points from 2.75 to 15.0 $\text{\AA}$:  with a step of 0.05 $\text{\AA}$ between 3.0 and 6.0 $\text{\AA}$, 0.10 $\text{\AA}$ between 6.0 and 9.5 $\text{\AA}$, and 2.50 $\text{\AA}$ from 10.0 to 15.0 $\text{\AA}$.
For the Sr–SrOH system, $\theta$ was varied from $0^\circ$ to $180^\circ$ in $10^\circ$ increments; for all other metal–SrOH complexes, potential cuts were computed at $\theta=0^\circ$, $90^\circ$, and $180^\circ$, corresponding to the metal–HOSr collinear, T‑shaped, and metal–SrOH collinear arrangements, respectively.
The norm of $T_1$ diagnostic, monitoring the reliability of the coupled cluster expansion, remained approximately 0.02 in all cases.\cite{Lee_1989}

The Sr–SrOH PES data points were expanded in Legendre polynomials as $V(R, \theta)=\Sigma_{\lambda=0}^{35} V_{\lambda}(R) P_{\lambda}(\cos\theta)$, where terms with $\lambda > 0$ represent anisotropic contributions to the interaction potential. To characterize the long‑range asymptotic region, supermolecular interaction energies with counterpoise correction were computed at the unrestricted coupled cluster singles, doubles and perturbative triples  [UCCSD(T)] level and projected onto the same Legendre basis. 
The long‑range van der Waals coefficients were then obtained by fitting the asymptotic behavior of the radial expansion coefficients $V_{\lambda}(R)$ to the UCCSD(T) PES. The long‑range PES was sampled in Jacobi coordinates with the radial coordinate 25.0 to 60.0 $\text{\AA}$ with a step of 2.50 $\text{\AA}$ between 25.0 and 30.0 $\text{\AA}$ and 5.00 $\text{\AA}$ between 30.0 and 50.0 $\text{\AA}$ at $\theta=0^\circ$ and  $180^\circ$ with $30^\circ$ increments.

\subsection{Field-free collisions of ${}^2 \Sigma$ molecules with ${}^1S$ atoms}
\label{sec:comp_details_collisions}

The collisional Hamiltonian was formulated to account for the center-of-mass motion, the interaction potential between the collision partners, and the rotational structure of the linear rigid molecule. 
The reduced mass, $\mu =47.83$ u, corresponding to ${}^{88}$Sr + ${}^{88}$Sr$^{16}$O$^{1}$H, was obtained from the NIST atomic masses~\cite{Emsley_1995}.
The rotational constant of SrOH was set at $B=0.249203$ cm${}^{-1}$, based on the experimental data~\cite{Steimle_1992}. 
The collision energy $E_{col}$ was fixed at 1 $\mu$K, ensuring the $s-$wave scattering regime, where only the single partial wave $L=0$ contributes.
The scattering length $a$ was extracted from the single‐channel probability amplitude and phase by solving the close-coupling equations~\cite{Arthurs_1960} using the hybrid log-derivative Airy propagator~\cite{Alexander_1984,Alexander_1987}.
The radial grid parameters for numerical propagation and the rotational basis‐set size truncation were optimized by monitoring the number of resonances in the scattering length spectrum as the interaction potential was scaled by $\gamma = 1 \pm 0.05$. 
This procedure yielded $R_{\mathrm{min}} = 2.5$ $\text{\AA}$, $R_{\mathrm{mid}}$ and $R_{\mathrm{match}} = 10.0$ $\text{\AA}$, $R_{\mathrm{max}} = 300.0$ $\text{\AA}$, and rotational states up to $j_{\mathrm{max}} = 50$. 
Finally, the same close-coupling methodology was applied to the bound-state problem, employing computational parameters identical to those used in the scattering length calculations.
All quantum scattering calculations were performed using \textsc{molscat} and \textsc{bound}~\cite{Hutson_2019_molscat, Hutson_2019b_bound}.

\subsection{Electronic structure of the excited states}
\label{sec:comp_details_excited_states}

Studies of excited states and transition dipole moments (TDMs) of the Sr–SrOH complex in the $C_s$ symmetry point group were investigated in two dissociation scenarios, both involving systematic variation of the Sr–O bond length while maintaining all other internal coordinates fixed. In the first case, the Sr–O distance was varied from the equilibrium value, with the remaining Sr–O–H geometry frozen at the fully optimized RCCSD(T) structure. In the second case, the Sr–O distance was similarly varied, but at the global minimum of the RCCSD(T) potential energy surface calculated in the Jacobi coordinate system.
The ground‐state potential energy curve was computed at the RCCSD(T) level. 
Excited‐state energies and TDM components were then obtained using the MCSCF level: three ${}^2A'$ and two ${}^2A''$ excited states were targeted, and the electric‐dipole transition moments from the $X^2\Sigma^+$ ground state to each excited state were evaluated directly. 
The vertical excitation energies were first determined as the differences between the ground- and excited‐state MCSCF energies at each geometry. To place these excited‐state curves on an absolute energy scale consistent with the ground state, each MCSCF excitation energy was shifted to the RCCSD(T) dissociation asymptote and added to the corresponding RCCSD(T) energy.

\label{sec:results}

\section{Electric Properties of the SrOH molecule}
\label{sec:results_properties}

The SrOH molecule, similarly to other heavy alkaline‐earth molecular radicals (CaF, SrF, CaOH, YbOH), exhibits a pronounced ionic bonding character. In the ground state $X^{2}\Sigma^+$, an alkaline‐earth metal atom transfers one of its two outer $ns^2$ electrons to the OH ligand, while the remaining electron occupies an open‐shell molecular orbital localized on the metal center. 
Specifically, the ground‐state orbital composition of linear SrOH is dominated by the Sr‑$s$ atomic orbital, as expected for an ionic radical~\cite{Li_2019}. This gives rise to small rotational constants and a dense rovibrational level structure with minimal vibrational branching~\cite{Lasner_2022}. 
The valence electron can be readily excited into higher orbitals, yielding nearly diagonal FCFs~\cite{Isaev_2016,Hao_2019,Wei_2025}. 

The permanent dipole moment ($\mu$) of SrOH, computed at the RCCSD(T) level, is 1.75 D, compared to the experimental value of 1.90 D~\cite{Steimle_1992}.  
At the RHF level, the result (0.97 D) significantly underestimates $\mu$ due to the neglect of electron correlation and the simplified treatment of ionic bonding.  
For comparison, CaOH has an experimental $\mu$ of 1.46 D~\cite{Steimle_1992}, SrF exhibits theoretical values of 3.46 D~\cite{Kosicki_2017,Hao_2019} and an experimental value of 3.47 D~\cite{Ernst_1985}, while CaF shows 3.07 D~\cite{Childs_1984}.  
Larger dipole moments correlate with increased molecular polarity, stronger intermolecular interactions, and more pronounced field orientation behavior.

At the RCCSD(T) level, the average polarizability ($\bar{\alpha}$) of SrOH is 182.11 a.u. with an anisotropy ($\Delta\alpha$) of 55.65 a.u.
By contrast, RHF yields 227.21 a.u. ($\bar{\alpha}$) and 108.06 a.u. ($\Delta\alpha$), again reflecting the absence of correlation effects.  
For further context, SrF at the CCSD(T) level has $\bar{\alpha} = 170.05$ a.u. and $\Delta\alpha = 66.35$ a.u.~\cite{Kosicki_2017}.  
Higher $\Delta\alpha$ indicates a stronger direction‑dependent response, which is particularly relevant for the manipulation of ultracold molecules. To the best of our knowledge, the polarizability of SrOH has not previously been reported. Now we may move to analyzing how SrOH interacts with other atoms.

\section{Interactions of SrOH with $s^2$-type atoms}
\label{sec:results_interactions}

\begin{table*}
\small
  \caption{\ Minimum energy $E_{\text{min}}$ (in cm$^{-1}$), dipole moment $\mu$ (in Debye), rotational constants A, B, C (in GHz), optimal geometries of alkaline-earth metal-$\text{SrOH}$,  $\text{Yb}$-$\text{SrOH}$, and $\text{Hg}$-$\text{SrOH}$ mixtures in the $X^2\Sigma^+$ electronic ground state. All bond lengths are in Angstroms (\r{A}) and bond angles are in degrees ($^\circ$).}
  \label{tbl:optg}
  \begin{tabular*}{\textwidth}{@{\extracolsep{\fill}}llllllllllll}
    \hline
    System & $E_{min}$  & $ \mu $& A & B & C& $r_{M-Sr}$ & $r_{M-O}$ & $r_{Sr-O}$ & $r_{O-H}$ & $\angle_{\text{M-O-Sr}}$ & $\angle_{\text{Sr-O-H}}$\\
    \hline
    \text{Mg} - \text{SrOH} & -6626  & 3.37 & 1.91 & 2.13 & 18.86 & 3.36 & 1.99 & 2.31 &  0.97 & 102 & 133 \\
    \text{Ca} - \text{SrOH} & -10565 & 1.82 & 1.28 & 1.38 & 16.23 & 3.60 & 2.21 & 2.33 &  0.97 & 105 & 132 \\
    \text{Sr} - \text{SrOH} & -10852 & 1.77 & 0.76 & 0.80 & 15.86 & 3.81 & 2.34 & 2.34 &  0.97 & 109 & 125 \\
    \text{Yb} - \text{SrOH} & -3483  & 1.74 & 0.53 & 0.55 & 12.26 & 3.92 & 2.79 & 2.23 &  0.97 & 102 & 143 \\
    \text{Hg} - \text{SrOH} & -629   & 1.23 & 0.58 & 0.62 & 8.08  & 3.55 & 3.36 & 2.13 &  0.97 & 77 & 178 \\
    \hline
  \end{tabular*}
\end{table*}

The equilibrium geometries of the electronic ground state optimized in the RCCSD(T) level for Mg-SrOH, Ca-SrOH, Sr${}_2$OH, Yb-SrOH, and at RMP2 for Hg-SrOH, are collected in Table~\ref{tbl:optg}. 
In the complexes examined, the M–O bond length increases systematically with the mass of M: from Mg (1.99 $\text{\AA}$)  to Sr (2.34 $\text{\AA}$), Yb (2.79 $\text{\AA}$) and further stretching to Hg (3.36 $\text{\AA}$).
In contrast, the Sr–O distance remains tightly clustered between 2.31 and 2.34 $\text{\AA}$ for Mg, Ca, Sr, and Yb complexes and slightly reduced for Hg (2.13 $\text{\AA}$). 
The O–H bond length is essentially constant (0.97 $\text{\AA}$), about 5\% and 1.5\% longer than the experimental values for SrOH (0.922 $\text{\AA}$) and CaOH (0.9562 $\text{\AA}$) respectively~\cite{Presunka_1995,  Kozyryev_2019}. 
This is consistent with the Sr-O, Mg-O, and O-H distances in isolated SrOH (2.16 and 0.97 $\text{\AA}$) and MgOH (1.76 and 0.95 $\text{\AA}$) obtained at RCCSD(T) level of theory~\cite{Li_2020, Koput_2023a}.
The M–O–Sr angle expands from $102^\circ$ (Mg) to $109^\circ$ (Sr) and then contracts for heavier metals (Yb $102^\circ$, Hg $77^\circ$), indicating increasing bond asymmetry. 
The Sr–O–H angle approaches linearity in the Hg complex ($178^\circ$), while it is strongly bent for Mg-SrOH, Ca-SrOH, and  Sr${}_2$OH ($133^\circ$, $132^\circ$, and $125^\circ$), consistent with earlier studies of the Li-SrOH system ($98.5^\circ$)~\cite{Morita_2017}.
For comparison, heteronuclear fluoride trimers exhibit even larger angular structures in their electronic ground state~\cite{Kosicki_2017}.
The correspondence between the fluoride complex angles and the OH complexes investigated here highlights how the choice of ligand governs bond anisotropy and overall molecular geometry.

The equilibrium energy $E_{min}$, rotational constants A, B, and C computed at the RCCSD(T) level, as well as the dipole moment $\mu$ obtained at the MCSCF level are also collected in Table~\ref{tbl:optg}.
The rotational constants $A$, $B$, and $C$ decrease systematically with heavier reduced masses. The lightest system, Mg–SrOH, displays the largest constants, whereas Sr${}_2$OH, Yb-SrOH, and Hg-SrOH exhibit near symmetric top constants with a ratio between $A/B$ near 99$\%$.
The deepest $E_{min}$ occur in  Sr${}_2$OH (–10852 cm${}^{-1}$) and Ca–SrOH (–10565 cm${}^{-1}$), while Yb-SrOH (–3483 cm${}^{-1}$) and Hg-SrOH are much shallower (–629 cm${}^{-1}$). 
Similarly, the dipole moment $\mu$ decreases from Mg (3.37 Debye) through Ca (1.82 Debye) and Sr (1.77 Debye) to Yb (1.74 Debye) and Hg (1.23 Debye). 
The lower $E_{min}$ and $\mu$ in complexes with heavier metals point to weaker binding and reduced polarizability of metals, likely due to the more diffuse electron shells of the heavier elements. 
In contrast, $E_{min}$ of –45047 cm${}^{-1}$ with $\mu$ = 3.46 Debye for the SrF molecule highlights the substantially stronger Sr–ligand bond in fluorides~\cite{Kosicki_2017}.
The relatively shallower $E_{\min}$ for Mg–SrOH (-6626 cm${}^{-1}$) than MgOH (-26200 cm${}^{-1}$)~\cite{Koput_2023a} demonstrates that the Mg–SrOH interaction is substantially weaker than the Mg–OH bond in the MgOH molecule. 
Likewise, the equilibrium energy of Sr${}_2$OH (–10852 cm${}^{-1}$) is much shallower than $E_{min}$ of the isolated SrOH molecule (-33300 cm${}^{-1}$)\cite{Li_2020}, reflecting the fact that the formation of the M–SrOH and Sr–SrOH complexes requires redistribution of electron density over two metal centers and the hydroxide ligand.
This delocalization leads to longer bonds and weaker binding compared to those of monohydroxide radicals.
The selection of metal atom (Mg, Ca, Sr, Yb, Hg) plays a decisive role in shaping both the geometry and the interaction strength of M–SrOH complexes. Depending on the selected metal, one can engineer either nearly symmetric top species such as Sr${}_2$OH, with a well-defined rotational structure and moderate dipole moment, or strongly asymmetric complexes, such as Mg–SrOH or Hg–SrOH, which offer enhanced anisotropy. This tunability opens opportunities to explore a diverse set of ultracold collision regimes, ranging from near-symmetric tops to highly anisotropic systems.

Despite these variations, all the species discussed here can be classified as nearly prolate symmetric tops, with Ray's $\kappa$ parameter from -0.97 to the extreme case of -0.997 for YbSrOH and Sr$_2$OH. Such near-symmetric tops exhibit features ideally suited for precision measurements and for exploring physics beyond the Standard Model: pairs of opposite parity energy levels that are very close in energy. Note also that the prolate near-symmetric tops experience a linear Stark effect in an electric field, which makes them highly tunable. Taking as an example the Sr$_2$OH system, we will now consider the possibility of forming them with association followed by STIRAP.

\begin{table}[h]
\small
  \caption{\ Relative energies $\Delta$E in (cm$^{-1}$) for the Sr-SrOH asymptotic dissociation channels.}
  \label{tbl:delta_E}
  \begin{tabular*}{0.48\textwidth}{@{\extracolsep{\fill}}ll}
    \hline
    Asymptotic channel & $\Delta$E  \\
    \hline
    \text{Sr} - \text{SrOH} & 0 \\
    \text{Sr$_2$} - \text{OH} & 54993.2  \\
    \text{SrO} - \text{SrH} & 59864.7  \\
    \hline
  \end{tabular*}
\end{table}

\section{Collisions of SrOH with Sr are non-reactive}
\label{sec:results_pes}

The data in Table~\ref{tbl:delta_E} demonstrate that both Sr${}_2$ + OH ($\Delta$E = 54993 cm$^{-1}$) and the SrO + SrH ($\Delta$E = 59865 cm$^{-1}$) channels lie far above the Sr + SrOH asymptote ($\Delta$E = 0 cm$^{-1}$).  As a result, Sr–SrOH is effectively non‐reactive under ultracold conditions: it cannot spontaneously dissociate into Sr${}_2$ + OH or SrO + SrH without supplying tens of thousands of wave numbers of energy.
It can be generalized to other alkaline-earth–SrOH systems, which distinguishes them from alkali-metal-SrOH collisions where chemical reactions lead to singlet and triplet products~\cite{Morita_2017}.
This pronounced endoergicity makes the Sr-SrOH system an ideal platform for controlled atom–molecule collisions, with the only possible chemical reaction being an isotope-substitution~\cite{Kosicki_2020a}. 
In fact, the isolated Sr${}_2$ molecule has a binding energy of only -1082 cm${}^{-1}$, an order of magnitude shallower than the Sr–SrOH well depth~\cite{Leung_2021}, while the exothermic energy of swapping a lighter isotope for a heavier strontium isotope in Sr${}_2$ is only about 0.003 cm${}^{-1}$~\cite{Tomza_2015}. These vast differences in energy scales mean that although Sr–SrOH remains inert to dissociation, a carefully prepared ultracold mixture can undergo isotope‐exchange chemistry with minimal energetic overhead, opening a clear path for precision studies of ultracold isotope effects.

\begin{figure*}
 \centering
\includegraphics[width=0.95\linewidth]{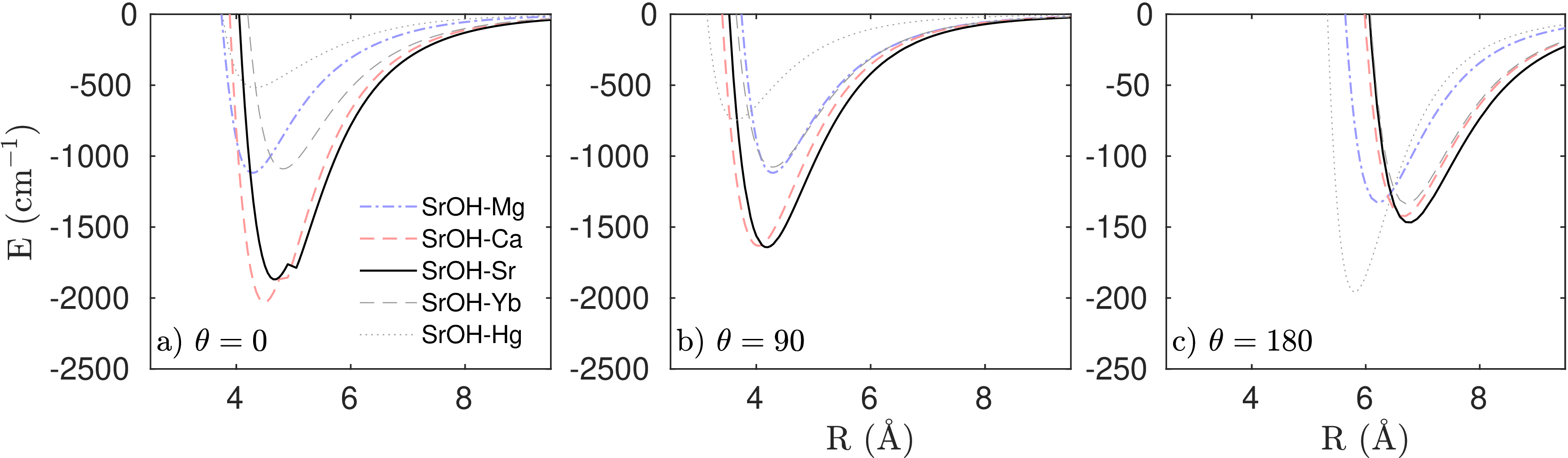}
\caption{Cuts through the potential energy surfaces of the $1^2A'$ ($X^2\Sigma^+$) electronic ground state of M–SrOH systems in Jacobi coordinates. Panel (a) shows the potential energy as a function of the Jacobi distance $R$ at the linear approach angle $\theta=0^\circ$, panel (b) at the T-shape geometry $\theta=90^\circ$, and panel (c) at the opposite linear geometry $\theta=180^\circ$. Solid lines correspond to M = Mg, Ca, and Sr; dashed lines to M = Yb; and the dotted line to M = Hg.} 
\label{fig:pes_cuts}
\end{figure*}

The computed one‐dimensional cuts through the PES with a fixed Jacobi angle of $\theta=0^\circ$, $90^\circ$, and $180^\circ$ for Mg-SrOH, Ca-SrOH, Sr${}_2$OH, Yb-SrOH, and Hg-SrOH complexes in the $1^2A'$ ($X^2\Sigma^+$) electronic ground state at the RCCSD(T) level of theory are presented in Figure~\ref{fig:pes_cuts}.
For linear ($0^\circ$) and T‐shaped ($90^\circ$) geometries, the depths and positions of the potential minima for Ca-SrOH, Sr-SrOH, Mg–SrOH, and Yb–SrOH are largely similar, although T-shaped configurations exhibit slightly shallower wells.
In opposite linear geometry ($180^\circ$), the Mg-SrOH, Ca-SrOH, Sr-SrOH, and Yb-SrOH systems exhibit a significant angular effect, with the potential minima shifting to larger Jacobi distance and well depths reduced by nearly an order of magnitude, yet still remaining above 150 cm${}^{-1}$.
In contrast, although Hg–SrOH exhibits the shallowest interaction potential among the systems studied, its well depth increases with the approach angle from $0^\circ$ to $90^\circ$, and at $\theta= 180^\circ$, it retains just over half of its depth.
The cuts through the PES are two orders of magnitude deeper than the He–CaH system in the $X^2\Sigma^+$ state~\cite{Balakrishnan_2003}, yet far more anisotropic.
Relative to Li–SrOH~\cite{Morita_2017}, where well depths of cuts also decrease with $\theta$, our results for $\theta=0^\circ$ and $90^\circ$ are nearly an order of magnitude deeper, while those for $\theta=180^\circ$ remain of comparable magnitude, further evidencing a stronger anisotropy. 
These features promise a rich ultracold collision dynamics with multiple resonances and strong rotational‐state mixing in studied complexes.

\begin{figure*}
 \centering
\includegraphics[width=0.7\linewidth]{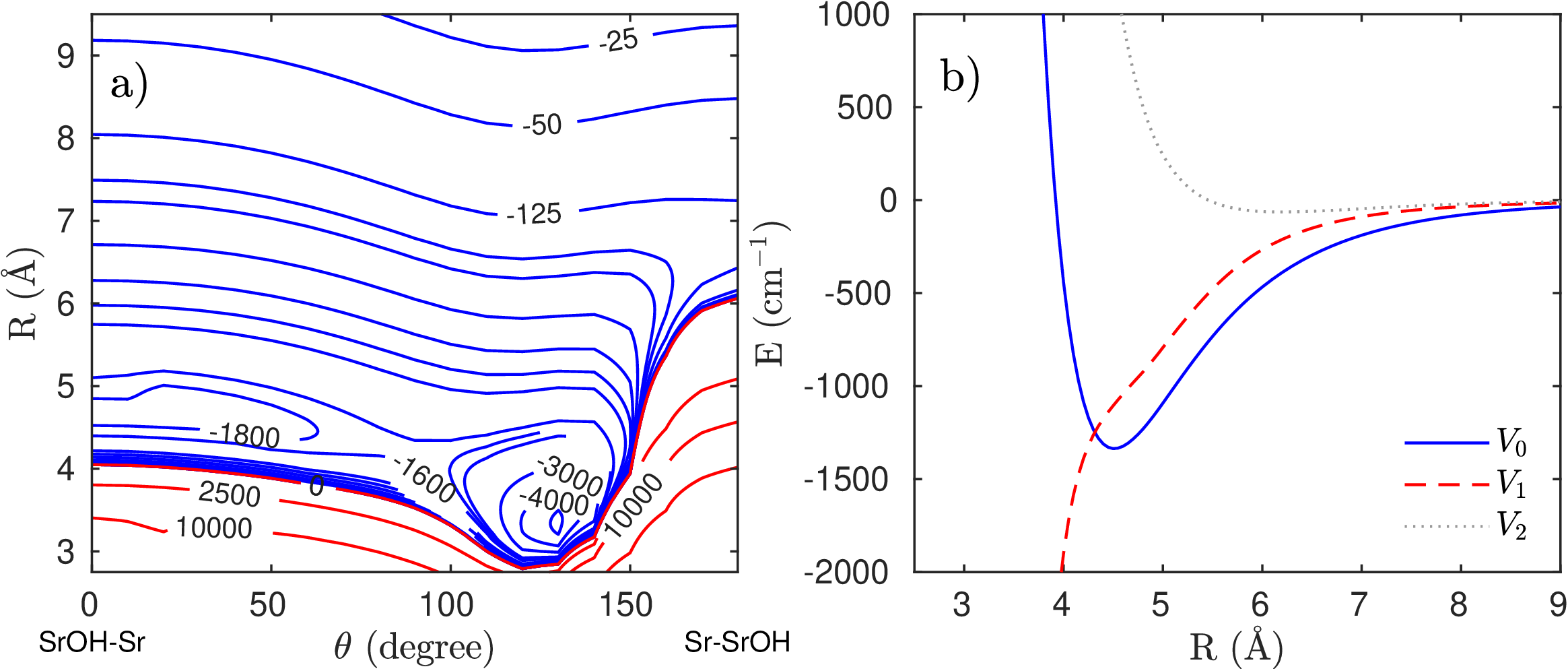}
 \caption{The interactions in the  $1^2A'$ ($X^2\Sigma^+$) electronic ground state of Sr–SrOH in the Jacobi coordinate system.
 Panel (a) shows the contour plot of the potential energy surface at the RCCSD(T) level of theory, as a function of Jacobi coordinates R and $\theta$ in units of cm$^{-1}$.
 Panel (b) shows the radial dependence of the first three Legendre coefficients $V_{\lambda}(R)$ $(\lambda=0, 1, 2)$ used in the expansion of the potential energy surface.
 }
 \label{fig:pes_and_expansion}
\end{figure*}

An inspection of Figure~\ref{fig:pes_and_expansion}(a) reveals that, although the  global minimum of Sr${}_2$OH occurs at -10852 cm${}^{-1}$, the potential energy surface expressed in the Jacobi coordinates exhibits a considerably shallower minimum of –5261 cm${}^{-1}$ at $R = 3.35$ $\text{\AA}$ and $\theta = 130^\circ$.
This difference reflects the fact that within the Jacobi representation, the SrOH molecule is constrained to a linear configuration and in the minimum relaxation of the whole complex leads to $C_{2v}$ symmetry.
A secondary stationary point is located at –1974 cm${}^{-1}$ ($R = 4.60$ $\text{\AA}$, $\theta = 40^\circ$), and two saddle points occur in linear geometries ($\theta = 0^\circ$ at $R = 4.65$ $\text{\AA}$, and $\theta = 180^\circ$ $R = 6.80$ $\text{\AA}$).
The pronounced variation in the interaction energy with $\theta$ underscores the strong anisotropy of the surface.
Analogous features, two minima and two saddle points, have been reported for the Sr${}_2$F (${1}^{2}A'$) surface, and the present Sr–SrOH PES is markedly more complex and anisotropic than those of high‑spin Li–SrOH~\cite{Morita_2017} or the isolated SrOH radical~\cite{Li_2020}.

Figure~\ref{fig:pes_and_expansion}(b) presents the radial dependence of the first three Legendre coefficients, $V_{\lambda}(R)$ $(\lambda = 0, 1, 2)$, for the potential energy surface of the $1^2A'$ ($X^2\Sigma^+$) ground state of the Sr–SrOH system, providing crucial insight into the anisotropy of the interaction.
The isotropic term $V_0(R)$ exhibits a well-defined minimum –1334 cm$^{-1}$ at ($R = 4.50$ $\text{\AA}$).
In contrast, the first anisotropic term, $V_1(R)$, shows no minimum and retains a significant magnitude over a wide range of distances, highlighting its dominant contribution to the interaction anisotropy, particularly in the short-range region. 
Higher-order anisotropic contributions exhibit the same repulsive character typical of short-range interactions as $V_2(R)$ and are therefore not shown here.
The observed Legendre coefficients show similarity to those found in the expansion of the potential energy surface of the high-spin Li–SrOH system~\cite{Morita_2017}.
Furthermore, fitting the asymptotic behavior of the Legendre coefficients yields long-range dispersion coefficients $C_{6,0}=3123$ and $C_{6,2}=50$ (both in atomic units), which quantify van der Waals interactions. Notably, the value of $C_{6,0}$
is more than $50\%$ lower than that reported for the Li–SrOH system~\cite{Morita_2017}, $19 \%$ lower than for Sr–SrF, and $9 \%$ lower than for Rb–SrF~\cite{Kosicki_2017}.

\section{Cold collisions of ground state SrOH with Sr atoms}
\label{sec:results_collisions}

\begin{figure*}
 \centering
\includegraphics[width=0.65\linewidth]{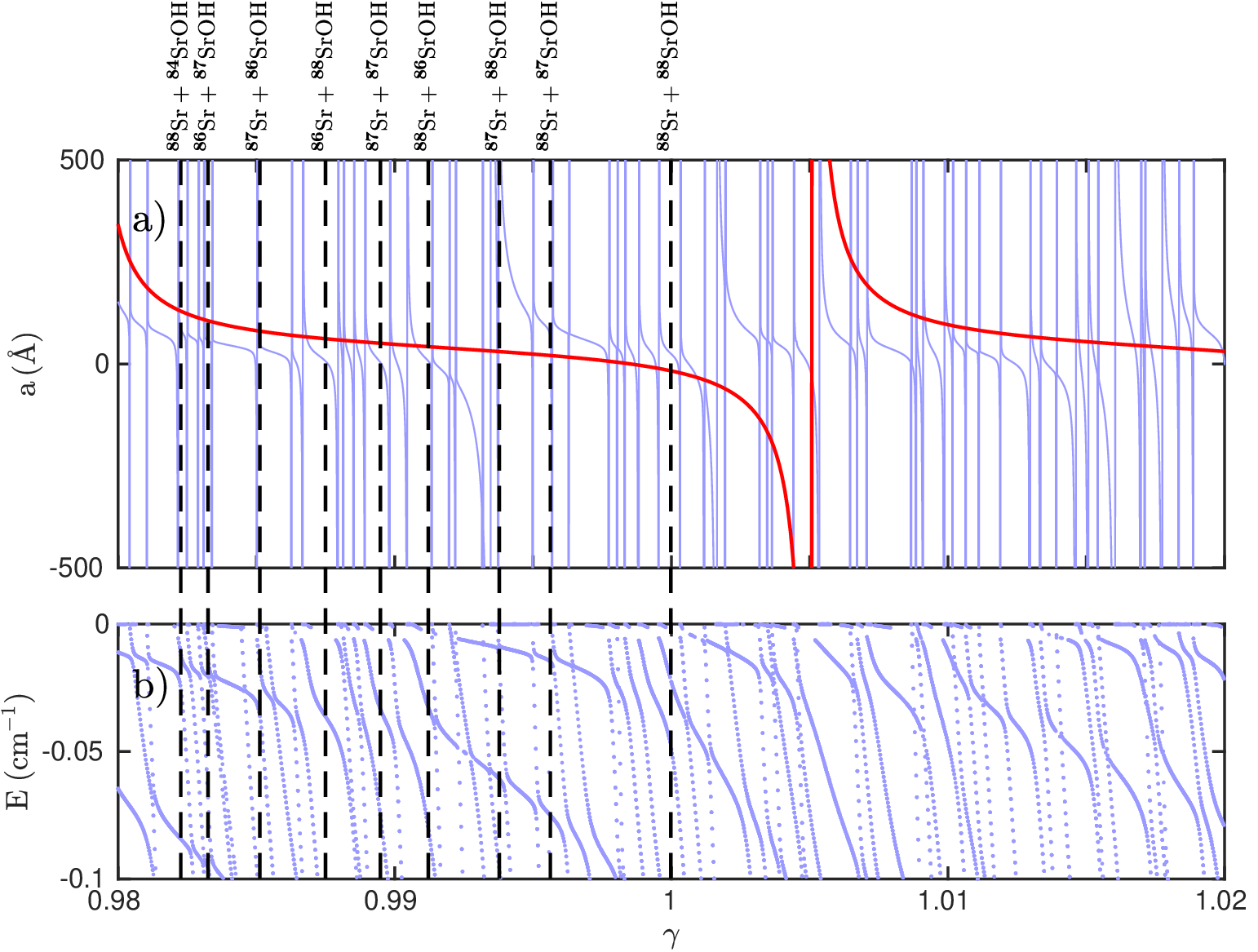}
 \caption{Scattering and near‐threshold bound‐state properties of the linear rigid‐rotor SrOH and Sr system in its electronic ground state $X{}^2\Sigma^+$ , as a function of potential‐energy scaling. Calculations are performed for s-wave collisions at collision energy $E_{col} = \mu\mathrm{K}$ and  rotational quantum number $j_{max} = 50$. Panel (a) shows the real part of the scattering length $a$ (in $\text{\AA}$) versus the scaling factor $\gamma$. The red curve corresponds to the isotropic part of the interaction potential only, while the thin light‐blue lines include the full Legendre‐expansion (isotropic + anisotropic terms) in Jacobi coordinates.
Panel (b) displays the near‐threshold bound‐state energies cm${}^{-1}$ as a function of the same scaling factor $\gamma$. Vertical dashed lines, common to both panels, mark the positions of resonances arising from different isotopic combinations of the Sr-SrOH complex
 }
 \label{fig:scatt}
\end{figure*}

The calculated isotropic dipolar polarizability of SrOH, 182 a.u. is comparable to that of atomic Sr ($\approx$197.14 a.u.\cite{Vitanov2017RMP}), suggesting that the Sr-SrOH pair is an excellent candidate for co-trapping in optical dipole traps or optical tweezers. In such an environment, one may pursue sympathetic cooling of SrOH from the tens of microkelvin down into the submicrokelvin regime. Furthermore, magnetically tunable Feshbach resonances could enable the association of Sr atoms with SrOH molecules into weakly bound complexes.   
Observation of such resonances requires the existence of sub-threshold molecular bound states that couple strongly to scattering channels. While a full prediction of the Feshbach spectrum for this highly anisotropic system would demand prohibitively large close-coupling basis sets and lies beyond our present scope, we can demonstrate 
very strong coupling between the ground and rotationally excited states of SrOH  using field-free atom+rigid rotor calculations of the scattering length with variable PES in which we simply apply mapping $V(R,\theta) \rightarrow \gamma V(R,\theta) $. Such a potential scaling to explore the density of states of the given system was used before by Hutson's group~\cite{Frye_2016}. Moreover,  by scaling the potential we can explore the sensitivity of scattering length calculations to uncertainty hidden in {\em ab intio } calculations, which is roughly about 10\%. 

In the range of potential‐energy scaling factors $\gamma$ between 0.95 and 1.05 the $s-$wave scattering length and the spectrum of near‐threshold bound states both exhibit extremely rich structure.
To facilitate a clear analysis, Figure ~\ref{fig:scatt} shows results only for the narrower interval $\gamma$ between 0.98 and 1.02, within which both spectra are very dense.
In panel (a), the thick red line shows the real part of the $s$ -wave scattering length calculated using only the isotropic component (Legendre term $V_{\lambda=0}$) of the interaction potential. This corresponds to the hypothetical case in which anisotropy is completely neglected. As is known from Gribakin and Flambaum's formula~\cite{Gribakin:1993} scattering length exhibits a pole whenever the bound state is exactly at threshold, thus a single broad resonance just above $\gamma=1.00$.

However, when the full anisotropic potential is included (thin light blue curves), the scattering length breaks up into a dense spectrum of narrow resonances associated with the rotational structure of the SrOH molecule. 
Each of these arises when a bound state is brought to the threshold by varying $\gamma$.   Generally, narrow resonances are produced by bound states with weaker coupling to the ground rotational state, in particular, to states dominated by higher rotational states $j$ and high end-over-end quantum number $L$, since $\mathbf{j} + \mathbf{L} = 0$.
Panel (b) confirms this observation: the near-threshold bound states (obtained with the same computational parameters as scattering length)  show numerous avoided crossings which are evidence of their coupling and actually cross zero energy at precisely the $\gamma$ values marked by resonances in the scattering length in panel (a). 
Vertical dashed lines, on both panels, indicate the positions of $\gamma$ uniquely associated with different isotopic combinations of Sr and SrOH, as labeled along the top of panel (a).
During the full $\pm 5\%$ scan in $\gamma$ we identify 159 distinct resonant features. This corresponds to $\sim 16$ resonances per $1 \%$ change in $\gamma$.
Such a high density of features underlines the extreme sensitivity of ultracold Sr-SrOH collisions to details of the short‐range potential, and illustrates the importance of including full anisotropic couplings when predicting or interpreting scattering properties in this system.

The spectrum of near-threshold bound states highlights the potential to form the Sr–SrOH complex by mergoassociation~\cite{Bird_2023, Ruttley_2023}.
By merging two optical tweezers~\cite{Anderegg_2019, Anderegg_2021, Bao_2023}, one containing a strontium atom and the other containing an SrOH molecule, slowly enough to follow the trap-induced avoided crossing between their separated motional state and a weakly bound Sr–SrOH level, one can adiabatically convert the pair into a molecule. 
Because Sr–SrOH exhibits a very dense spectrum of bound states, there will be multiple closely spaced avoided crossings; adiabatic passage across the most strongly coupled crossing may convert the atom–fragment pair into a bound Sr–SrOH molecule.

Figure~\ref{fig:scatt}  clearly demonstrates that the source of resonances in the Sr-SrOH system is the anisotropy of the potential energy surface. 
The observed resonance spectrum as a function of potential scaling is significantly more congested than in our previous study of ErYb~\cite{Kosicki_2020b}.
In systems containing highly magnetic lanthanide atoms~\cite{Frisch_2014, Frisch_2015,Maier_2015a, Augustovicova_2018b, Kosicki_2020b}, the interplay between strong interaction potential anisotropy and Zeeman coupling under an external magnetic field gives rise to a chaotic resonance distribution, as evidenced by level spacings following the Wigner-Dyson distribution~\cite{Maier_2015a}.
Similarly, the Sr–SrOH system may exhibit a chaotic resonance structure, since its rotational degree of freedom effectively serves as orbital angular momentum in highly magnetic systems.
Indeed, Li-CaH and Li-CaF resonance spectra display chaotic signatures in both the nearest-neighbor spacing distributions and level number variance~\cite{Frye_2016}. 
Although a comprehensive statistical analysis of resonance widths and mean spacings in Sr-SrOH is beyond the scope of this work, these preliminary observations motivate further investigations into the detailed nature of its resonance spectrum.

\section{Pathway toward ground-state Sr$_2$OH} 
\label{sec:results_pathway}

\subsection{Reducing complexity of the problem}

Having discussed the dense spectrum of bound states in the Sr–SrOH system, we can reasonably assume that, when both species are prepared in their absolute ground states, weakly bound complexes can be produced either by magnetic field ramps or by merging optical traps. In this section, we examine whether such complexes can be converted into deeply bound Sr–SrOH molecules, in particular, the absolute (rovibronic) ground state.

The preparation of deeply bound molecules in the electronic ground state typically relies on STIRAP via suitably chosen electronically excited intermediate states of the dimer. The key figures of merit are the Rabi frequencies for the pump and Stokes transitions, which are set by the transition dipole moments and the Franck–Condon overlaps between the initial, intermediate, and target levels. For polyatomic systems, assessing STIRAP feasibility is challenging primarily because accurate FCFs)are hard to obtain: full-dimensional wavefunctions are generally unavailable beyond triatomic molecules (and even triatomics remain demanding).

However, as shown below in this section, the ground state and a set of low-lying excited states are largely parallel, with no complications such as crossings or avoided crossings. We therefore implement a simple reduced-dimensionality model in which the SrOH geometry and the Sr–O–H angle are held fixed while only the separation between the second Sr atom and the SrOH unit is varied. We consider two limiting cases for Sr detachment: (i) a linear SrOH molecule with the incoming Sr approaching from infinity, and (ii) the Sr$_2$OH complex in its equilibrium geometry from which Sr detaches (see below for details). We emphasize that the goal of the present study is to obtain order-of-magnitude estimates: how many STIRAP steps are required, how deep in binding one can proceed in a single step, and how strong the transition-dipole couplings are between the ground and intermediate states.

The second approximation introduced here is the neglect of spin–orbit interactions. We focus on states correlating with the first excited states of both the Sr atom and the SrOH molecule. The lowest excited state of strontium, $^3P$, is spin-forbidden, while the $^2\Pi$ state of SrOH is allowed and notably strong. In our case, it is sufficient to consider the properties arising from the behavior of the valence electrons. Although spin–orbit coupling may shift energy levels and render some formally forbidden transitions weakly allowed, the overall character of the low-lying excited states is primarily governed by the non-relativistic Hamiltonian.

\subsection{Low lying excited states of Sr-SrOH}
\label{sec:results_excited_states}

\begin{figure*}
\includegraphics[width=0.7\linewidth]{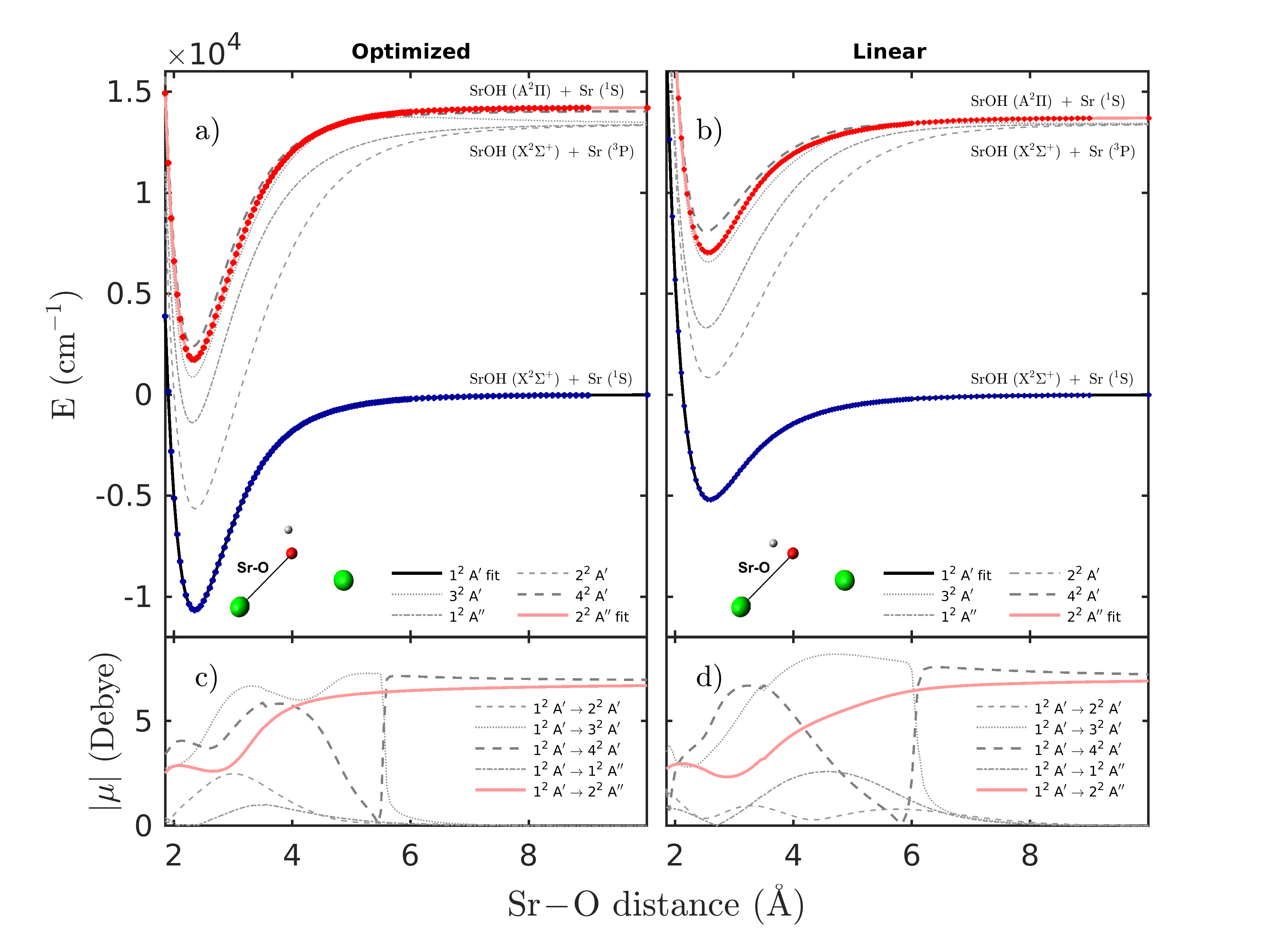}
 \caption{
One-dimensional cuts through the Sr$_2$OH potential energy and transition dipole-moment surfaces along a single Sr–O bond coordinate.
Panels (a) and (c) correspond to the {\it Optimized} geometry with all other coordinates fixed at the RCCSD(T) equilibrium of the $1^2A'$ ground state, while panels (b) and (d) correspond to the {\it Linear} geometry with the SrOH fragment constrained to linearity.
Panels (a) and (b) show potential energy curves of the ground and low-lying excited doublet states, and panels (c) and (d) present the corresponding transition dipole moments. The insets illustrate the stretched Sr–O bond.}
 \label{fig:pec_tdm}
\end{figure*}

The Sr–SrOH asymptotes can be addressed using the same lasers employed for laser cooling and trapping. For SrOH, one drives strong allowed transitions, while the narrow $^3P_1$ line in Sr has nevertheless been used successfully for photoassociation. One could therefore envision a two-step transfer: first, a detuned laser excites into an SrOH($^2\Pi$)–dominated state, then a second photon drives the system back down to the electronic ground state.

Panels (a) and (b) of Figure \ref{fig:pec_tdm} display the potential energy curves for the ground and excited states as functions of the Sr–O bond length, whereas panels (c) and (d) show the corresponding transition dipole moments from the 1${}^2A'$ ground state.
We present two Sr–O bond‐length scans to probe the potential energy and transition dipole moment surfaces, referred to as the \textit{Optimized} and \textit{Linear} cases.
In the \textit{Optimized} case, a single Sr–O bond is stretched while all remaining internal coordinates of the complex remain fixed at the RCCSD(T) equilibrium geometry of the $1^2A'$ ground state (Table \ref{tbl:optg}).
In the \textit{Linear} case, the Sr–O bond is varied within the fixed Jacobi‐coordinate geometry corresponding to the RCCSD(T) minimum of the $1^2A'$ state, where the SrOH fragment dissociating from Sr is fixed to a linear arrangement. 
Insets in the lower right of panels (a) and (b) illustrate the two scanning protocols.

Figure~\ref{fig:pec_tdm} (a) reveals that all potential energy curves for the ground ($1^2A'$) and all excited ($2^2A'$, $3^2A'$, $4^2A'$, $1^2A''$, $2^2A''$) states show a smooth behavior with well-defined minima occurring at nearly the same Sr-O bond length. In the asymptotic limit, they correlate to clear distinct dissociation channels: SrOH($X^2\Sigma^+$)+Sr($^1S$) for $1^2A'$, SrOH($X^2\Sigma^+$)+Sr($^3P$) for $1^2A''$, $2^2A'$, $3^2A'$, and SrOH(A$^2\Pi$)+Sr($^1S$) for $2^2A''$, $4^2A'$.
The ground-state well depth –10852 cm${}^{-1}$ at $2.34$ $\text{\AA}$) agrees with the optimized geometry result reported in Table~\ref{tbl:optg}.
The $2^2A'$ and $1^2A''$ excited states exhibit wells of –5648 cm${}^{-1}$ at $2.35$ $\text{\AA}$ and –1394 cm${}^{-1}$ at $2.30$ $\text{\AA}$, respectively, and all remaining wells lie above the SrOH($X^2\Sigma^+$)+Sr($^1S$) dissociation threshold.
The $4^2A'$ and $2^2A''$ states intersect near $4.73$ $\text{\AA}$, which is allowed due to their different symmetries ($A'$ and $A''$, respectively), and thus do not repel each other. 
A conical intersection occurs between the $3^2A'$ and $4^2A'$ states, both of $A'$ symmetry, around $5.50$ $\text{\AA}$, beyond which the $3^2A'$  state clearly correlates with SrOH($X^2\Sigma^+$)+Sr($^3P$) dissociation limit.
In this region, the $3^2A'$ and $4^2A'$ states converge energetically within 200 cm$^{-1}$ of the $2^2 A''$ state, indicating strong mixing between these manifolds. 
Such mixing can both facilitate and complicate population transfer schemes for STIRAP purposes, depending on the timing and spectral bandwidth of the laser pulses. 
Furthermore, the nearly parallel shapes of these PECs imply similar vibrational level spacings, suggesting an enhancement of Franck–Condon overlaps for selected transitions.

Comparing the one‑dimensional cuts in Figure ~\ref{fig:pec_tdm} (a) and (b), we find that the overall shapes of the ground and excited‐state potential energy curves and their asymptotic correlations remain essentially the same, despite modest shifts in well depths and Sr-O bond lengths induced by the strong anisotropy of the Sr–SrOH interaction. 
In the second scan over the Sr–O bond stretched within fixed the Jacobi‑coordinate geometry corresponding to the computed energy minimum ( ~\ref{fig:pec_tdm} (b)), the ground‐state well is shallower (-5188 cm${}^{-1}$ at $2.60$ $\text{\AA}$) and all excited‐state minima appear slightly farther out ($2.50-2.55$ $\text{\AA}$), but each curve still correlates cleanly to the same dissociation channels SrOH($X^2\Sigma^+$)+Sr($^1S$) for $1^2A'$, SrOH($X^2\Sigma^+$)+Sr($^3P$) for $1^2A''$, $2^2A'$, $3^2A'$, and SrOH(A$^2\Pi$)+Sr($^1S$) for $2^2A''$, $4^2A'$.
The symmetry‐allowed crossings of $4^2A'$–$2^2A''$ ($5.95$ $\text{\AA}$) and the conical intersection of $3^2A'$–$4^2A'$ ($6.10$ $\text{\AA}$) occur likewise at essentially the same Sr-O distance, with even stronger near‐degeneracy (about 30 cm${}^{-1}$) reflecting robust manifold mixing. 

An inspection of Figure~\ref{fig:pec_tdm} (c) shows the absolute values of the TDMs from the $1^2A'$ ground state to the respective excited states, computed at the MCSCF level.  
Clear distinctions are evident in the behavior of TDM as a function of the Sr–O distance.
For states $1^2A''$, $2^2A'$, and $3^2A'$, asymptotically correlating to SrOH($X^2\Sigma^+$)+Sr($^3P$), the TDMs decrease sharply beyond Sr-O bond length beyond $6.50$ $\text{\AA}$.
Such decay is characteristic of transitions that become forbidden in the atom–molecule asymptote or reflect weakening orbital overlap.
In contrast, the TDMs for the $2^2A''$ and $4^2A'$ states, correlating to SrOH(A$^2\Pi$)+Sr($^1S$),
remain above 6 Debye even at a large Sr-O distance. 
This persistent coupling suggests that these transitions can be exploited in coherent control schemes such as STIRAP to achieve efficient population transfer without the need for short‑range collisions.

Likewise, the transition dipole moments (Figure ~\ref{fig:pec_tdm} (d)) exhibit the same qualitative behavior: rapid decay beyond the $6.50$ $\text{\AA}$ Sr-O distance for states correlated with Sr($^3P$) and persistently large (higher than 6 Debye) values for transitions into SrOH(A$^2\Pi$)+Sr($^1S$), confirming that favorable Franck–Condon overlaps and strong long‑range coupling crucial for STIRAP persist in linear geometry. 
Although the numerical shifts in well depths and bond lengths will require minor adjustments to laser wavelengths and pulse bandwidths, they do not undermine the overall feasibility of implementing a one‑dimensional STIRAP scheme in Sr–SrOH.

At short-range the excited Sr and SrOH states strongly mix, as evidenced by the rapid variation of their transition dipole moments (see the previous section). Moreover, the potential curves for the lowest few excited states are nearly parallel for both geometries we study, which most likely makes the FCFs very favorable. Even when the Sr atom detaches from Sr–OH (varying the Sr–O–H angle), the ground and first excited states of Sr$_2$OH remain essentially the same single electronic configuration.

For an efficient STIRAP transfer, one must achieve sufficiently large Rabi frequencies linking the initial, intermediate, and final states. These Rabi frequencies scale with the Franck–Condon overlaps and the optical transition dipoles. In our tetrameric Sr$_2$OH complex, the strong dipole moments of SrOH and Sr asymptotes mix in the short-range, guaranteeing large transition moments toward excitation near either asymptote.

\begin{figure*}
\centering
\includegraphics[width = 0.8\linewidth]{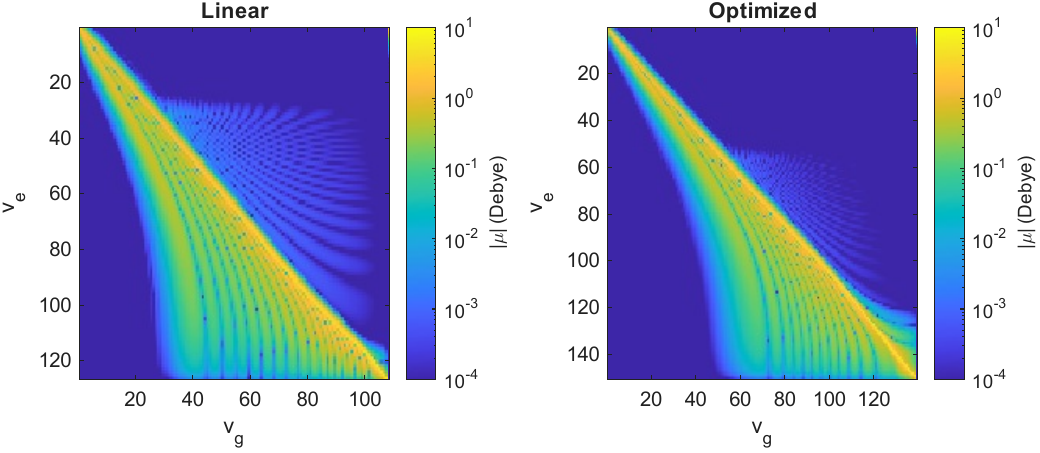}
\caption{Transition dipole moments between vibrational levels $v_g$ in the ground 1~$^2$A' state and $v_e$ in the excited 2~$^2$A'' state, calculated in both the \emph{linear} and \emph{optimized} geometries (see text for details).\label{fig:moments_map}}
\end{figure*}

\subsection{Choosing the intermediate state for STIRAP}
\label{sec:intermediate_state}

A successful STIRAP sequence must bridge the gap between the large internuclear separations explored by the initial near-threshold bound state and the short equilibrium distance where the rovibrational state resides. This requires choosing an excited bound state that satisfies two conditions. First, the vibrational wavefunction must have some overlap with the wavefunctions of both the initial and the target states. Second, the electronic transition dipole moment $\mu(R)$ must be appreciable at all the internuclear distances involved. Of all the calculated excited potential curves, only the $2\,^{2}A''$ state supports a consistently non-zero transition dipole moment spanning all internuclear separations.

To find out if there are viable STIRAP pathways for the association of Sr$_2$OH we will, for simplicity, ignore the potential anisotropy. This will allow us to use a single-channel radial equation 
\begin{equation}
\left(-\frac{\hbar^2}{2\mu}\frac{d^2}{dR^2} + V_{g,e}(R)\right) \Psi_{g,e}^v(R)  = E_{g,e}^v \Psi(R)_{g,e}^v(R)
\end{equation}
for both the (electronic) ground- ($g$) or excited- ($e$) states to obtain bound state energies $E_{g,e}^v$ and wavefunctions $\Psi_{g,e}^v(R)$. Throughout this section, we will use a reduced mass $\mu$ that corresponds to the $^{88}$Sr+$^{88}$Sr$^{16}$O$^{1}$H isotopologue. Then, we will calculate the transition dipole moments between individual vibrational states $v_g$ and $v_e$. However, since we need the potential curves to extend to much larger distances $R$ than our \emph{ab initio} data, we use a fitted analytic model instead.

Since Sr-SrOH is a van der Waals system, it is permissible to model the interaction potential with a Tang-Toennies formula. We choose the following variant\cite{Patkowski2009Be2}:
$V(R) = \left(A+BR+C/R+DR^2+ER^3\right) \times e^{-bR} - f_6(bR)\frac{C_6}{R^6}$ where $f_{n}$ is a Tang-Toennies damping function\cite{Tang1984}. We fit the parameters of the analytic model to the \emph{ab initio} data using the least-squares method.

In our fits we vary only the short-range polynomial parameters $A$ through $E$, while keeping the Born-Mayer range parameter $b=0.7$ constant. To fix the long-range van der Waals parameter $C_6$ we employ a combination rule: we start with the van der Waals coefficient of Sr$_2$, $C_6(\rm{Sr}_2) = 3164.3\,{\rm a.u.}$\cite{Stein2010Sr2} and scale it by the ratio of scalar dc polarizabilities of the Sr atom $\alpha(\rm{Sr}) = 197.14\,{\rm a.u.}$ \cite{Safronova2013BBR} and the SrOH monomer $\alpha(\rm{SrOH}) = 182.11\,{\rm a.u.}$
 $   C_6(\rm{Sr-SrOH}) \approx C_6(\rm{Sr}_2) \times \alpha(SrOH)/\alpha(Sr) = 2923.1\,{\rm a.u.}$

\subsection{Theoretical spectroscopy and STIRAP pathways}
\label{sec:results_stirap}

A key quantity that determines the strength of a transition is its transition dipole moment; it will determine the achievable Rabi frequencies when the transition is interrogated by a laser. To evaluate it for a transition between bound states labeled $v_g$ (ground electronic state) and $v_e$ (excited electronic state), we numerically calculate the matrix element,
\begin{equation}
    \mu(v_g, v_e) = \int_{0}^{\infty} \Psi_g^{v_g}(R) \mu(R) \Psi_e^{v_e} (R) dR
\end{equation}
where $\mu(R)$ is the $R$-dependent \emph{ab initio} transition dipole moment reported above. These are shown as a function of the vibrational quantum numbers in Figure~\ref{fig:moments_map}.

Even though the two models have differing potential depths and support different number of vibrational levels -- the linear model supports 108 and 126 bound states in the ground and excited states, while the optimized model supports 139 and 150 -- overall the behavior of the transition dipole moments is similar in both models. 

\begin{figure}
\centering
\includegraphics[width = 0.9\columnwidth]{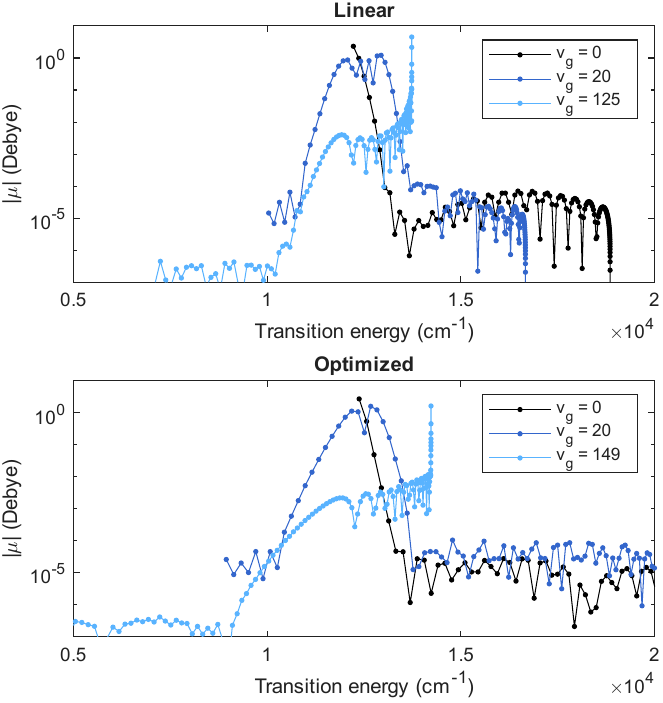}
\caption{Transition dipole moments for selected initial ground-state vibrational levels as a function of transition energy. Most of the strong transitions lie in the near-infrared-to-visible range.\label{fig:moments_vs_energy}}
\end{figure}

Next, we find out how the optical spectra of Sr$_2$OH could vary depending on the initial ground-electronic-state vibrational level. In Figure~\ref{fig:moments_vs_energy}, we show transition moments as a function of the transition energy for several selected initial states. The alternate $x$ axis shows the transition wavelength.

Most of the spectroscopic activity lies in the near-infrared to visible region. Starting from the near-threshold state, many transitions in the 10000 cm$^{-1}$ to about 14000 cm$^{-1}$. Similarly, when starting from a deeply bound state $v_g=20$, there is a broad intensity maximum from about 10500 cm$^{-1}$ to about 14000 cm$^{-1}$. Transitions between weakly bound states depend on the SrOH excitation energy, but useful transitions could go as deep as 9000 cm$^{-1}$.

\begin{figure}
\centering
\includegraphics[width = 0.9\columnwidth]{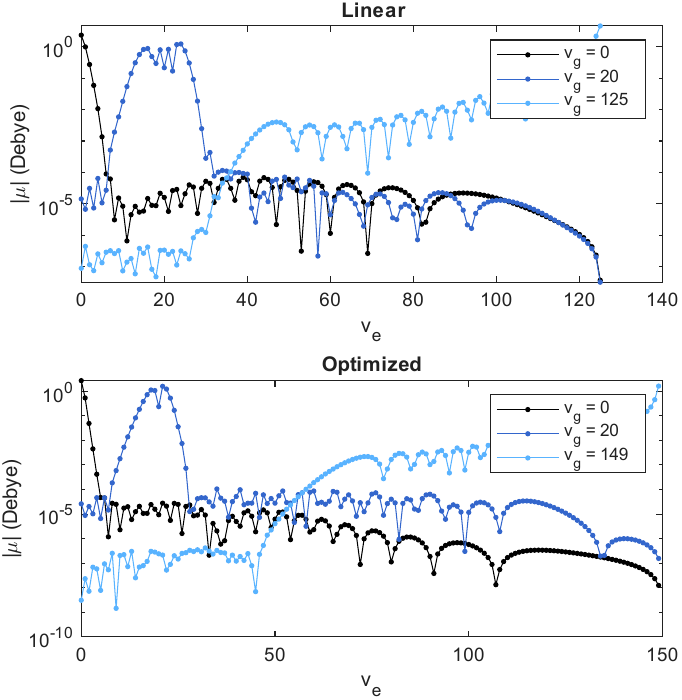}
\caption{Transition dipole moments for selected initial ground-state vibrational levels as a function of excited vibrational level. \label{fig:moments_vs_state}}
\end{figure}

Last, we consider the possibility of using STIRAP~\cite{Vitanov2017RMP} to transfer a Feshbach-associated weakly bound Sr$_2$ OH molecule to rovibrational ground state. Because the ground- and excited-state potentials are so similar, the FCFs are nearly diagonal. This will likely preclude finding viable three-level (two-photon) STIRAP pathways. However, five-state (four-photon) paths could exist, as suggested by the "linear" geometry model. For example, one could start from a weakly bound state and transition to the $v_g = 20$ state via the $v_e = 30$ excited state. Then, one could reach the Sr$_2$OH rovibrational ground state $v_g = 0$ via $v_e = 6$.

\section{Summary and conclusions}
\label{sec:summary}

Motivated by the growing interest in cooling and trapping polyatomic molecules, we have carried out a theoretical investigation of the interactions between SrOH and alkaline-earth–like atoms (e.g., Mg, Ca, Sr, Yb, Hg). These systems are central to state-of-the-art precision-spectroscopy experiments; controlled assembly by linking SrOH with an alkaline-earth metal atom via electromagnetic fields could open access to a new class of molecules with interesting and structural properties, like near symmetric tops. By analogy to alkali-metal dimers, such species may also serve as platforms for studies of ultracold chemical reactions and few-body dynamics.

First, we characterized the ground-state electric properties of SrOH at the RCCSD(T) level, obtaining a permanent dipole moment of 1.75 D as well as the average polarizability of 182.11 a.u. reported here for the first time. Second, we optimized the geometries and determined the binding energies of SrOH complexes with several $s^2$-type atoms (Mg, Ca, Sr, Yb, Hg).  The results indicate that lighter metals form more polar molecules, whereas heavier metals exhibit stronger anisotropic interactions of potential energy surfaces.
All studied ground-state tetratomic molecules are near symmetric tops with anisotropy parameter at least -0.97 and should behave much like symmetric tops.
In particular, the Sr${}_2$SrOH complex shows a well depth of –10852 cm$^{-1}$ and a dipole moment of 1.77 D, making it suitable for both stable trapping via electric field. 
Third, a detailed study of the potential energy surface showed that the Sr-SrOH system is non-reactive under ultracold conditions, with all possible chemical channels lying tens of thousands of cm$^{-1}$ above the entrance channel. 
The PES is strongly anisotropic, supporting deep wells and exhibiting a large variation with the Jacobi angle. Similar properties were found for all systems except Hg-SrOH, which is much shallower and much less anisotropic. Legendre expansion analysis confirmed that the first anisotropic term $V_1(R)$ dominates the angular dependence in the short-range. Fourth, quantum scattering calculations using scaled interaction potentials demonstrated that anisotropy gives rise to an exceptionally dense resonance spectrum. The high resonance density arises from the involvement of numerous rotationally excited levels of SrOH and the strong anisotropy of the interactions.

Finally, we investigated pathways for a two steps optical association of Sr$_2$OH in its rovibrational ground state via STIRAP. 
The low-lying excited states of the complex were computed for two geometries with streched Sr-O, revealing nearly parallel potential curves and large transition dipole moments for states correlating to SrOH(A$^2\Pi$)+Sr($^1S$) hindered by near-diagonal Franck–Condon factors sequences offer a viable route to deep molecular binding.

In conclusion, the Sr–SrOH system combines favorable trapping and stability properties with strong and tunable anisotropy, enabling both precise control of ultracold collisions and the potential creation of asymmetric top molecules in the ground state. The detailed PESs, scattering predictions, and excited-state characterizations reported here provide a quantitative foundation for upcoming experimental efforts in this direction.

\section*{Acknowledgments}

M.B.K. and M.U., and P.S.\.Z. were supported by the National Science Centre, Poland (grant no. 2020/36/C/ST4/00508 and 2021/41/B/ST2/00681, respectively).
We are grateful to Wroclaw Centre for Networking and Supercomputing (grant no. 218) for a generous amount of CPU time.
MB gratefully acknowledges Poland's high-performance Infrastructure PLGrid PCSS for providing computer facilities and support within computational grant no. pl0516.

\section*{Author contributions}
Electronic structure calculations were primarily performed by M.B.K.
Scattering length calculations were conducted by M.B.K. with assistance from P.S.Z.
Bound-state calculations were carried out by M.B.K. with contributions from M.U. and P.S.Z.
STIRAP calculations were performed by M.B.
M.B.K., M.B., and P.S.Z. jointly contributed to the preparation and editing of the manuscript.

\section*{Conflicts of interest}
There are no conflicts to declare.

\section*{Data availability}
All data supporting the findings of this study, including electronic structure results and quantum scattering calculations, are openly available in the GitHub repository at \href{https://github.com/maciejkosicki/Production-of-ultracold-asymmetric-tops-from-Sr-atoms-and-SrOH-molecules.git}{https://github.com/maciejkosicki/Production-of-ultracold-asymmetric-tops-from-Sr-atoms-and-SrOH-molecules.git}.



\balance



\bibliographystyle{rsc}
\providecommand{\noopsort}[1]{}\providecommand{\singleletter}[1]{#1}%
\providecommand*{\mcitethebibliography}{\thebibliography}
\csname @ifundefined\endcsname{endmcitethebibliography}
{\let\endmcitethebibliography\endthebibliography}{}

\end{document}